\documentclass[twocolumn,showpacs,preprintnumbers,amsmath,amssymb]{revtex4}
\usepackage{graphicx}
\usepackage{dcolumn}
\usepackage{epsfig,graphicx,amsmath}
\begin{document}

\title{Harmonic entanglement in a degenerate parametric down conversion}
\author{Sintayehu Tesfa}
 \email{sint_tesfa@yahoo.com}
\affiliation{Physics Department, Addis Ababa University, P. O. Box 1176, Addis Ababa, Ethiopia}%


\date{\today}

\begin{abstract}We study the harmonic entanglement and squeezing in a two-mode
radiation produced in a degenerate parametric down conversion
process coupled to a two-mode vacuum reservoir employing the
linearization procedure. It is found that there is a quadrature
entanglement between the harmonically related fundamental and
second-harmonic modes and the superimposed radiation exhibits a
significant two-mode squeezing. The entanglement exits even when
there is no two-mode squeezing, since the correlation leading to
these phenomena are essentially different. In addition, the more
the external coherent light is down converted, the more stronger
the entanglement and mean photon number of the two-mode radiation
would be which is not generally true for squeezing.
\end{abstract}

 \maketitle

\section{INTRODUCTION}

Optical degenerate parametric down convrsion is one of the
second-order nonlinear processes in which a pump photon of
frequency $2\omega$ is down converted into a pair of signal
photons each of frequency $\omega$. Due to the inherent two-photon
nature of the interaction, the parametric oscillator or amplifier
is one of the most interesting and well studied devices in the
nonlinear quantum optics
\cite{1,2,3,4,5,6,7,sint071,sint072,sintq}. It is found to be a
good source of light with nonclassical features. The quantum
optical properties are significantly degraded by the leakage
through the mirrors and amplification of the quantum fluctuations
in the cavity. However, if the ordinary vacuum is replaced by a
squeezed vacuum reservoir, the nonclassical properties can be
enhanced \cite{6,7,8} provided that the reservoir is squeezed in
the right quadrature, since the fluctuations entering the cavity
is biased. Moreover, when a nonlinear crystal is shined with an
external radiation of frequency $2\omega$ only some part of this
radiation is down converted into a pair of photons \cite{sint071}.
As a result, the cavity contains the down converted and unchanged
radiations. Although a considerable attention has been given to a
single-mode squeezed radiation previously, most recently it is
envisioned that a degenerate parametric oscillator can be a source
of a two-mode light with nonclassical properties characterized by
a strong correlation between the states of each mode separately as
well as the superimposed state formed by the two radiations
\cite{sint072}.

Currently, generation of macroscopic entangled states is receiving
attention in connection with their potential in the quantum
information and measurement theories \cite{9,10}. At microscopic
level nonclassical correlations and entanglement arise due to the
down conversion of a single high frequency photon into a pair of
correlated lower frequency photons. The generation of the
entangled beams in the nondegenerate parametric down conversion
was predicted by Reid and Drummond \cite{11,12,13}. Studies have
also shown the existence of the correlation between the
fundamental and second-harmonic modes \cite{14} including
entanglement in the second-harmonic generation \cite{15}. In this
respect, Lim and Saffman \cite{16} analyzed the production of two
beams with nonclassical intensity correlations and quadrature
entanglement in the dual-ported reservoir of the second-harmonic
generation and found that the harmonic output exhibits strong
quantum correlations. In connection to this, Grosse {\it{et al.}}
\cite{17} predicted a perfect entanglement between the fundamental
and second-harmonic modes in the pump depleted nondegenerate
parametric amplification and defined such an entanglement between
harmonically related fields as harmonic entanglement.

Einstein-Podolsky-Rosen (EPR) presented in 1935 \cite{18} their
famous argument, which on the basis of the local realism claimed
that an observation of perfectly correlated positions and momenta
would imply the incompleteness of the quantum theory. The
essential step in realizing the EPR-type entanglement is then to
introduce correlated states of at least two particles that persist
even when the particles are spatially separated. As thoroughly
discussed by Dechoum {\it{et al.}} \cite{19} without having to
construct an experimentally impossible states with perfect
correlation, as in the original paper, it is possible to
demonstrate this type of correlation using the inferred Heisenberg
principle. In this regard, a direct and experimentally feasible
qualitative criterion for such a correlation of continuous
variables was first proposed by Reid \cite{12} via quadrature
phase amplitude, applying a nondegenerate parametric
amplification, which is closely related to the original version.
In addition, making use of the quadrature variables Lodahl
\cite{lod} showed the existence of EPR-type correlations in the
second-harmonic generation. Most recently, Olsen \cite{15} has
considered a travelling-wave
 second-harmonic generation and
 predicted that this quantum system can be employed in the experimental demonstration
of entanglement with continuous variables. It then appears natural
to ask whether the strong correlation between the fundamental and
second-harmonic modes of the degenerate parametric oscillator
associated with the down conversion also leads to entanglement or
not? The main task of this communication is, therefore, devoted to
answer this question.

In this paper, the nonclassical properties of the two-mode cavity
radiation of the driven degenerate parametric oscillator coupled
to a two-mode vacuum reservoir applying the linearization
procedure in which the quantum properties of the system in time
are taken to vary slightly around the steady state mean values
would be analyzed. As discussed by Gilles {\it{et al.}} \cite{21},
this approximation remains valid as long as the quantum
fluctuations are much smaller than the classical mean values which
is the essence of a weaker coupling of the radiation with the
oscillator that corresponds to small quadrature noise. We also
consider the semiclassical approximation whereby the two modes are
assumed to be uncorrelated at steady state. As recently reported
by Chaturvedi {\it{et al.}} \cite{22}, the semiclassical theory is
found to work surprisingly well in the threshold region.
 We hence study the harmonic entanglement and its relation with the squeezing
 near threshold, since the relation between the two is an interesting
 issue by its own right \cite{prl93063601,prl95120502}.
 Moreover, we investigate the efficiency with which a nonlinear crystal down convert the
 light falling on it and the association of the down conversion with the entanglement
 and squeezing. We also calculate the mean photon number to see how intense
 the generated light could be.
%

\section{Quantum Langevin Equations}

Interaction of an external coherent radiation with a nonlinear
crystal responsible for a degenerate parametric oscillation placed
in a resonant cavity  can be described in the rotating-wave
approximation and in interaction picture by the Hamiltonian of the
form
\begin{align}\label{dp02}\hat{H}_{I}={i\lambda\over2}
\left[\hat{a}^{\dagger^{2}}\hat{b}-\hat{a}^{2}\hat{b}^{\dagger}\right]
+ i\varepsilon\left[\hat{b}^{\dagger} - \hat{b}\right],\end{align}
where $\varepsilon$ is proportional to the amplitude of the
coherent input, $\lambda$ is the measure of the coupling of a
nonlinear crystal with the external coherent radiation, $\hat{a}$
and $\hat{b}$ are the time-independent annihilation operators for
the fundamental (subharmonic) and pumping (second-harmonic) modes,
and $\lambda$ and $\varepsilon$ are chosen to be real-positive
constants. In view of the fact that $\hat{a}$ and $\hat{b}$ are
mutually commuting operators, the pertinent quantum Langevin
equations are found to be
\begin{align}\label{dp03}{d\hat{a}\over dt}
=\lambda\hat{a}^{\dagger}\hat{b}
-{\kappa\over2}\hat{a}+\hat{F}_{a}(t),\end{align}
\begin{align}\label{dp04}{d\hat{b}\over dt} = -{\lambda\over2}\hat{a}^{2}
 -{\kappa\over2}\hat{b}+\varepsilon+\hat{F}_{b}(t),\end{align}
where $\kappa$ is the cavity damping constant chosen to be the
same for both modes and $\hat{F}_{i}(t)$, with $i=a,b$, are the
Langevin noise operators satisfying, for a two-mode vacuum
reservoir, the correlation functions:
\begin{align}\label{dp05}\langle\hat{F}_{i}(t)\rangle=0,\end{align}
\begin{align}\label{dp06}\langle\hat{F}_{i}^{\dagger}(t)\hat{F}_{j}(t')\rangle=
\langle\hat{F}_{i}^{\dagger}(t)\hat{F}_{j}^{\dagger}(t')\rangle=\langle\hat{F}_{i}(t)\hat{F}_{j}(t')\rangle
=0,\end{align}
\begin{align}\label{dp07}\langle\hat{F}_{i}(t)\hat{F}_{i}^{\dagger}(t')\rangle
=\kappa\delta(t-t'),\end{align}
\begin{align}\label{dp08}\langle\hat{F}_{i}(t)
\hat{F}^{\dagger}_{j}(t')\rangle_{i\neq j}=0.\end{align}

We notice that Eqs. \eqref{dp03} and \eqref{dp04} are nonlinear
coupled differential equations that can be solved employing the
linearization procedure. In this approach, we first take
\begin{align}\label{dp09}\hat{a}(t) = \alpha+\hat{A}(t),\end{align}
\begin{align}\label{dp10}\hat{b}(t) = \beta+\hat{B}(t),\end{align}
where $\hat{A}(t)$ and $\hat{B}(t)$ are very small variations
about the mean values at steady state, and we define
\;\;$\alpha=\langle\hat{a}(t)\rangle_{ss}$ and
$\beta=\langle\hat{b}(t)\rangle_{ss}$. This approximation remains
valid as long as the quantum fluctuations about the mean values
are much smaller than the classical mean values, which implies
that the quantum noise during the interaction is quite small. Upon
taking the statistical average of Eqs. \eqref{dp03} and
\eqref{dp04} and then using the semiclassical approximation,
whereby at steady state the modes are assumed to be uncorrelated,
$\langle\hat{a}^{\dagger}(t)\hat{b}(t)\rangle_{ss}=
\langle\hat{a}^{\dagger}(t)\rangle_{ss}\langle\hat{b}(t)\rangle_{ss},$
and classical decorrelation that operators are assumed to be
factorized,
$\langle\hat{a}^{2}(t)\rangle_{ss}=\langle\hat{a}(t)\rangle^{2}_{ss},$
we obtain
\begin{align}\label{dp11}\lambda\alpha^{*}\beta
-{\kappa\over2}\alpha=0,\end{align}
\begin{align}\label{dp12}\lambda\alpha^{2}
+\kappa\beta=2\varepsilon.\end{align} We realize that the
semiclassical assumption is found to work for weak nonlinearity or
weak coupling between the external radiation and nonlinear crystal
where the mean photon number at threshold is very large.

Now with the aid of Eqs. \eqref{dp03}, \eqref{dp04}, \eqref{dp09},
\eqref{dp10}, \eqref{dp11}, \eqref{dp12}, and the fact that
$\hat{A}$ and $\hat{B}$ are small perturbations, it is possible to
verify that
\begin{align}\label{dp13}{d\hat{A}(t)\over
dt}=\varepsilon_{1}^{*}\hat{B}(t)
+\varepsilon_{2}\hat{A}^{\dagger}(t)-{\kappa\over2}\hat{A}(t)+\hat{F}_{a}(t),\end{align}
\begin{align}\label{dp14}{d\hat{B}(t)\over
dt}=-\varepsilon_{1}\hat{A}(t)
-{\kappa\over2}\hat{B}(t)+\hat{F}_{b}(t),\end{align} where we set
$\varepsilon_{1}^{*}=\lambda\alpha^{*}$ and
$\varepsilon_{2}=\lambda\beta$. Multiplication of Eq. \eqref{dp11}
by $\alpha$ suggests that the phase of $\beta$ is the same as that
of $\alpha^{2}$. This is consistent with Eq. \eqref{dp12} if both
$\alpha$ and $\beta$ are real, which implies that
$\varepsilon_{1}$ and $\varepsilon_{2}$ are also real.
 It then follows from Eqs.
\eqref{dp11} and \eqref{dp12} that
\begin{align}\label{dp15}\varepsilon_{1}=\pm\sqrt{2\lambda\varepsilon-\kappa\varepsilon_{2}}
\end{align} and $\varepsilon_{2}=\kappa/2$ for
$\varepsilon_{1}\ne0$. We notice that there are two possible
values for the fundamental amplitude and hence the system may be
in a transient superimposed state of these amplitudes prior to
detection \cite{10}.
%
Moreover, the solutions of these coupled differential equations
are found following a straight forward algebra to be
\begin{align}\label{dp16}\hat{A}(t) &=
a_{1}(t)+\big[a_{3}(t)+pa_{4}(t)\big]
\hat{A}(0)+\big[a_{5}(t)+pa_{6}(t)\big]\notag\\&\times
\hat{A}^{\dagger}(0)
+qa_{6}(t)\hat{B}(0)+qa_{4}(t)\hat{B}^{\dagger}(0)+\hat{f}(t),\end{align}
\begin{align}\label{dp17}\hat{B}(t) &=
a_{2}(t)+\big[a_{3}(t)-pa_{4}(t)\big]
\hat{B}(0)+\big[a_{5}(t)-pa_{6}(t)\big]\notag\\&\times\hat{B}^{\dagger}(0)
-qa_{6}(t)\hat{A}(0)-qa_{4}(t)\hat{A}^{\dagger}(0)+\hat{g}(t),\end{align}
where
\begin{align}\label{dp18}\hat{f}(t) &= \int_{0}^{t}\big[\big(a_{3}(t-t')+pa_{4}(t-t')\big)
\hat{F}_{a}(t')
\notag\\&+\big(a_{5}(t-t')+pa_{6}(t-t')\big)\hat{F}^{\dagger}_{a}(t')
\notag\\&+qa_{6}(t-t')\hat{F}_{b}(t')+qa_{4}(t-t')\hat{F}^{\dagger}_{b}(t')\big]dt',\end{align}
\begin{align}\label{dp19}\hat{g}(t) &= \int_{0}^{t}\big[\big(a_{3}(t-t')-pa_{4}(t-t')\big)
\hat{F}_{b}(t')
\notag\\&+\big(a_{5}(t-t')-pa_{6}(t-t')\big)\hat{F}^{\dagger}_{b}(t')
\notag\\&-qa_{6}(t-t')\hat{F}_{a}(t')-qa_{4}(t-t')\hat{F}^{\dagger}_{a}(t')\big]dt',\end{align}
in which
\begin{align}\label{dp20}p={\varepsilon_{2}\over\sqrt{\varepsilon_{2}^{2}-4\varepsilon_{1}^{2}}},
\end{align}
\begin{align}\label{dp21}q={2\varepsilon_{1}
\over\sqrt{\varepsilon_{2}^{2}-4\varepsilon_{1}^{2}}},
\end{align}
\begin{align}\label{dp22}a_{1}(t)&=-{e^{-{(\kappa-\varepsilon_{2})
t\over2}}\over\lambda}\left[\varepsilon_{1}\cosh\left({t\sqrt{\varepsilon_{2}^{2}
-4\varepsilon_{1}^{2}}\over2}\right)\right.\notag\\&\left.+\big(\varepsilon_{1}p+\varepsilon_{2}q\big)
\sinh\left({t\sqrt{\varepsilon_{2}^{2}
-4\varepsilon_{1}^{2}}\over2}\right)\right],\end{align}
\begin{align}\label{dp23}a_{2}(t)&=-{e^{-{(\kappa-\varepsilon_{2})
t\over2}}\over\lambda}\left[\big(\varepsilon_{1}q+\varepsilon_{2}p\big)
\sinh\left({t\sqrt{\varepsilon_{2}^{2}
-4\varepsilon_{1}^{2}}\over2}\right)\right.\notag\\&\left.-\varepsilon_{2}
\cosh\left({t\sqrt{\varepsilon_{2}^{2}
-4\varepsilon_{1}^{2}}\over2}\right)\right],\end{align}
\begin{align}\label{dp24}a_{3}(t)=e^{-{\kappa
t\over2}}\cosh\left({\varepsilon_{2}t\over2}\right)\cosh\left({t\sqrt{\varepsilon_{2}^{2}
-4\varepsilon_{1}^{2}}\over2}\right),\end{align}
\begin{align}\label{dp25}a_{4}(t)=e^{-{\kappa
t\over2}}\sinh\left({\varepsilon_{2}t\over2}\right)\sinh\left({t\sqrt{\varepsilon_{2}^{2}
-4\varepsilon_{1}^{2}}\over2}\right),\end{align}
\begin{align}\label{dp26}a_{5}(t)=e^{-{\kappa
t\over2}}\sinh\left({\varepsilon_{2}t\over2}\right)\cosh\left({t\sqrt{\varepsilon_{2}^{2}
-4\varepsilon_{1}^{2}}\over2}\right),\end{align}
\begin{align}\label{dp27}a_{6}(t)=e^{-{\kappa
t\over2}}\cosh\left({\varepsilon_{2}t\over2}\right)\sinh\left({t\sqrt{\varepsilon_{2}^{2}
-4\varepsilon_{1}^{2}}\over2}\right).\end{align} It may worth to
mention that Eqs. \eqref{dp16} and \eqref{dp17} are applied to
calculate various quantities of interest. We also observe that the
condition for $\varepsilon_{1}$ to be real requires that
$2\lambda\varepsilon\ge\kappa\varepsilon_{2}$. We, therefore,
denote the case for which
$2\lambda\varepsilon=\kappa\varepsilon_{2}$ that corresponds to
$\varepsilon_{1}=0$ as a threshold condition.

\section{Quadrature Entanglement}

In this section, we seek to study the entanglement of the
fundamental and second-harmonic modes in the cavity. It is a
well-established fact that a quantum system is said to be
entangled, if it is not separable. That is, if the density
operator for the combined state cannot be expressed as a
combination of the product density operators of the constituents,
\begin{align}\label{dp28}\hat{\rho}\ne\sum_{j}\hat{\rho}_{j}^{(1)}\bigotimes\hat{\rho}_{j}^{(2)}.\end{align}
On the other hand, entangled continuous variable state can be
expressed as a co-eigenstate of a pair of EPR-type operators
 such as $\hat{X}_{a}-\hat{X}_{b}$ and
$\hat{P}_{a}+\hat{P}_{b}$ \cite{pra74}. The total variance of
these two operators reduces to zero for maximally entangled
continuous variable states. Nonetheless, according to the
criterion set by Duan {\it{et al.}} \cite{duan} quantum states of
the system are entangled, provided that the sum of the variances
of a pair of EPR-like operators
\begin{align}\label{dp29}\hat{u}=\hat{X}_{a}-\hat{X}_{b},\end{align}
\begin{align}\label{dp30}\hat{v}=\hat{P}_{a}+\hat{P}_{b},\end{align}
where
$\hat{X}_{a}={1\over\sqrt{2}}\big(\hat{a}+\hat{a}^{\dagger}\big)$,
$
\hat{X}_{b}={1\over\sqrt{2}}\big(\hat{b}+\hat{b}^{\dagger}\big)$,
\newline$
\hat{P}_{a}={i\over\sqrt{2}}\big(\hat{a}^{\dagger}-\hat{a}\big)$,
and
$\hat{P}_{b}={i\over\sqrt{2}}\big(\hat{b}^{\dagger}-\hat{b}\big)$,
 satisfy
\begin{align}\label{dp31}\Delta u^{2}+\Delta v^{2}<2,\end{align}
in which
\begin{align}\label{dp32}\Delta u^{2}+\Delta v^{2}&=2\langle\hat{a}^{\dagger}\hat{a}\rangle+
2\langle\hat{b}^{\dagger}\hat{b}\rangle+2\langle\hat{a}\rangle\langle\hat{b}\rangle+
2\langle\hat{a}^{\dagger}\rangle\langle\hat{b}^{\dagger}\rangle\notag\\&-
2\langle\hat{a}\hat{b}\rangle-2\langle\hat{a}^{\dagger}\hat{b}^{\dagger}\rangle
-2\langle\hat{a}^{\dagger}\rangle\langle\hat{a}\rangle-
2\langle\hat{b}^{\dagger}\rangle\langle\hat{b}\rangle
+2.\end{align}

Next we determine the various correlations in Eq. \eqref{dp32}. To
this end, assuming the cavity mode to be initially in the vacuum
state and using the fact that the Langevin noise forces have zero
mean along with Eqs. \eqref{dp09}, \eqref{dp10}, \eqref{dp16}, and
\eqref{dp17}, we get
\begin{align}\label{dp33}\langle\hat{a}(t)\rangle=\alpha
+a_{1}(t),\end{align}
\begin{align}\label{dp34}\langle\hat{b}(t)\rangle=\beta
+a_{2}(t),\end{align}
\begin{align}\label{dp35}\langle\hat{a}^{\dagger}(t)\hat{a}(t)\rangle=\big(\alpha+a_{1}(t)\big)^{2}
+\langle\hat{f}^{\dagger}(t)\hat{f}(t)\rangle,\end{align}
\begin{align}\label{dp36}\langle\hat{b}^{\dagger}(t)\hat{b}(t)\rangle=\big(\beta+a_{2}(t)\big)^{2}
+\langle\hat{g}^{\dagger}(t)\hat{g}(t)\rangle,\end{align}
\begin{align}\label{dp37}\langle\hat{a}(t)\hat{b}(t)\rangle=(\alpha+a_{1}(t))(\beta+a_{2}(t))
+\langle\hat{f}(t)\hat{g}(t)\rangle.\end{align} Moreover, in view
of the correlations of the Langevin noise forces \eqref{dp06},
\eqref{dp07}, and \eqref{dp08}, it is possible to show at steady
state that
\begin{align}\label{dp38}\langle\hat{f}^{\dagger}(t)\hat{f}(t)\rangle_{ss} &={\kappa[(1+p^{2}
+q^{2})(\kappa-\varepsilon_{2})-2
p\sqrt{\varepsilon_{2}^{2}-4\varepsilon_{1}^{2}}]
\over8[\kappa(\kappa-2\varepsilon_{2})+4\varepsilon_{1}^{2}]}
 \notag\\&+
{\kappa[(1+p^{2} +q^{2})(\kappa+\varepsilon_{2})+2
p\sqrt{\varepsilon_{2}^{2}-4\varepsilon_{1}^{2}}]\over8[\kappa(\kappa+2\varepsilon_{2})
+4\varepsilon_{1}^{2}]} \notag\\&+{\kappa^{2}(1-p^{2}
-q^{2})\over4(\kappa^{2}-\varepsilon_{2}^{2})}-{\kappa^{2}(1-p^{2}+q^{2})\over
4[\kappa^{2}-(\varepsilon_{2}^{2}-4\varepsilon_{1}^{2})]}\notag\\&-{(1+p^{2}-q^{2})\over4},\end{align}
\begin{align}\label{dp39}\langle\hat{g}^{\dagger}(t)\hat{g}(t)\rangle_{ss} &={\kappa[(1+p^{2}
+q^{2})(\kappa-\varepsilon_{2})+2p\sqrt{\varepsilon_{2}^{2}-4\varepsilon_{1}^{2}}]\over8[\kappa(\kappa-2\varepsilon_{2})+4\varepsilon_{1}^{2}]}
\notag\\&+ {\kappa[(1+p^{2}
+q^{2})(\kappa+\varepsilon_{2})-2p\sqrt{\varepsilon_{2}^{2}-4\varepsilon_{1}^{2}}]
\over8[\kappa(\kappa+2\varepsilon_{2})+4\varepsilon_{1}^{2}]}
\notag\\&+{\kappa^{2}(1-p^{2}
-q^{2})\over4(\kappa^{2}-\varepsilon_{2}^{2})}
-{\kappa^{2}(1-p^{2}+q^{2})\over
4[\kappa^{2}-(\varepsilon_{2}^{2}-4\varepsilon_{1}^{2})]}\notag\\&-{(1+p^{2}-q^{2})\over4},\end{align}
\begin{align}\label{dp40}\langle\hat{f}(t)\hat{g}(t)\rangle_{ss} &=-{\kappa pq
(\kappa-\varepsilon_{2})\over4[\kappa(\kappa-2\varepsilon_{2})+4\varepsilon_{1}^{2}]}
\notag\\&- {\kappa
qp(\kappa+\varepsilon_{2})\over4[\kappa(\kappa+2\varepsilon_{2})+4\varepsilon_{1}^{2}]}
+{qp\kappa^{2}\over2(\kappa^{2}-\varepsilon_{2}^{2})}\notag\\&-{\kappa^{2}pq\over
2[\kappa^{2}-(\varepsilon_{2}^{2}-4\varepsilon_{1}^{2})]}+{pq\over2}.\end{align}
We, therefore, see with the aid of Eqs. \eqref{dp32},
\eqref{dp33}, \eqref{dp34}, \eqref{dp35}, \eqref{dp36},
\eqref{dp37}, \eqref{dp38}, \eqref{dp39}, and \eqref{dp40} at
steady state that
\begin{align}\label{dp41}\Delta u^{2}+\Delta v^{2} &=2+{\kappa(1+p^{2}
+q^{2}+2qp)(\kappa-\varepsilon_{2})\over2[\kappa(\kappa-2\varepsilon_{2})+4\varepsilon_{1}^{2}]}
\notag\\&+ {\kappa(1+p^{2}
+q^{2}+2qp)(\kappa+\varepsilon_{2})\over2[\kappa(\kappa+2\varepsilon_{2})+4\varepsilon_{1}^{2}]}
\notag\\&+{\kappa^{2}(1-p^{2}
-q^{2}-2qp)\over(\kappa^{2}-\varepsilon_{2}^{2})}\notag\\&-{\kappa^{2}(1-p^{2}+q^{2}-2qp)\over
[\kappa^{2}-(\varepsilon_{2}^{2}-4\varepsilon_{1}^{2})]}\notag\\&-(1+p^{2}-q^{2}+2qp).\end{align}
Finally, on account of Eqs. \eqref{dp20}, \eqref{dp21}, and the
fact that $\varepsilon_{2}=\kappa/2$ for $\varepsilon_{1}\ne0$, we
find
\begin{align}\label{dp42}\Delta u^{2}+\Delta v^{2} &={\kappa^{3}(\kappa+4\varepsilon_{1})
\over8\varepsilon_{1}^{2}(\kappa^{2}-16\varepsilon_{1}^{2})}\notag\\&+{3\kappa^{3}
(\kappa+4\varepsilon_{1})\over4(\kappa^{2}-16\varepsilon_{1}^{2})(\kappa^{2}+2\varepsilon_{1}^{2})}
\notag\\&-{16\varepsilon_{1}(4\varepsilon_{1}
+\kappa)+24\kappa\varepsilon_{1}\over3(\kappa^{2}-16\varepsilon_{1}^{2})}
\notag\\&+{16\kappa^{3}\varepsilon_{1}\over(\kappa^{2}-16\varepsilon_{1}^{2})(3\kappa^{2}
+16\varepsilon_{1}^{2})}.\end{align} It is not difficult to
observe that $\Delta u^{2}+\Delta v^{2}$ is very large when
$\varepsilon_{1}=0$ and $\varepsilon_{1}=0.25\kappa$. We hence
plot $\Delta u^{2}+\Delta v^{2}$ versus $\varepsilon_{1}$ for
$\varepsilon_{1}>0.25\kappa$.

\begin{figure}[hbt]
\centerline{\includegraphics [height=6cm,angle=0]{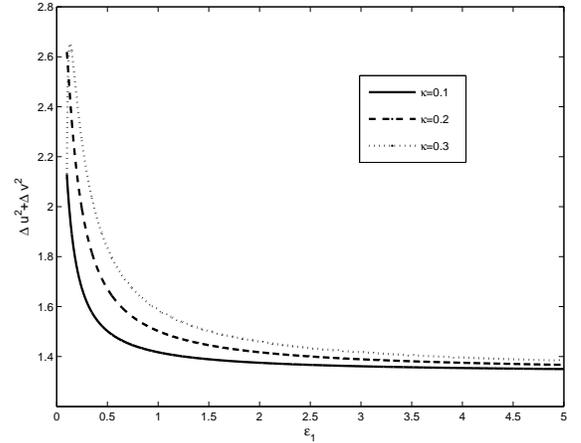} }
\caption {Plots of the sum of the variances of the EPR-type
operators  ($\Delta u^{2}+\Delta v^{2}$) at steady state for
different values of $\kappa$.}
\end{figure}

As clearly shown in Fig. 1 the correlation between the fundamental
and second-harmonic modes exhibits a quadrature entanglement
except near certain value of $\varepsilon_{1}$, for example
$\varepsilon_{1}=0.13$ for $\kappa=0.3$ at steady state. We also
realize that entanglement does not exist near a threshold value,
$\varepsilon_{1}=0$. However, entanglement exists and it decreases
with damping constant in other cases for which
$\varepsilon_{1}>0.25\kappa$. This is related to a well known fact
that the lesser the cavity damping constant, the more the
radiation stays in the cavity which in turn enhances the
correlation that leads to entanglement. It is possible to observe
that the dependence of the entanglement on damping through the
mirrors is insignificant for larger values of $\varepsilon_{1}$.
Moreover, as can easily be seen from Eq. \eqref{dp32} the existing
entanglement is attributed to the correlation between the states
of the two cavity modes. We realize that since
$\varepsilon_{1}=\lambda\langle\hat{a}\rangle_{ss}$, it
corresponds to the degree at which the external radiation of
frequency $2\omega$ is down converted by the nonlinear crystal.
Therefore, it can be inferred from Fig. 1 that the entanglement
would be stronger, the more efficiently the external radiation is
down converted by the crystal. This indicates that even though the
down conversion process breaks the coherent external radiation
into two, it is unable to destroy the coherence that is
responsible for the correlation between the down converted and
unchanged radiations.
%
%
%

\section{Quadrature variances}

In this section, we seek to analyze the squeezing properties of
the two-mode cavity radiation which can be described by
annihilation operator
\begin{align}\label{dp43}\hat{c}={1\over\sqrt{2}}\big(\hat{a}+\hat{b}\big),\end{align}
where $\hat{a}$ and $\hat{b}$ are the boson operators that
represent the fundamental and second-harmonic modes. In view of
the boson commutation relation for $\hat{a}$ and $\hat{b}$, one
can easily see that $\big[\hat{c},\;\hat{c}^{\dagger}\big]=1$ and
$\big[\hat{c},\;\hat{c}\big]=0$. We note that the squeezing of the
two-mode cavity radiation can be studied using the quadrature
operators corresponding to $\hat{c}$,
\begin{align}\label{dp46}\hat{c}_{+}=\hat{c}^{\dagger}+\hat{c}\end{align}
and
\begin{align}\label{dp47}\hat{c}_{-}=i\big(\hat{c}^{\dagger}-\hat{c}\big),\end{align}
in which the squeezing occurs when
\begin{align}\label{en48}\Delta c_{\pm}^{2}&=1+\langle\hat{a}^{\dagger}\hat{a}\rangle+
\langle\hat{b}^{\dagger}\hat{b}\rangle\pm(\langle\hat{a}\hat{b}\rangle
+\langle\hat{a}^{\dagger}\hat{b}^{\dagger}\rangle)+\langle\hat{a}^{\dagger}\hat{b}\rangle
\notag\\&+\langle\hat{a}\hat{b}^{\dagger}\rangle\pm{1\over2}\big[
\langle\hat{a}^{\dagger^{2}}\rangle+\langle\hat{b}^{\dagger^{2}}\rangle
+\langle\hat{a}^{2}\rangle+\langle\hat{b}^{2}\rangle+\langle\hat{a}\rangle^{2}
\notag\\&+\langle\hat{b}\rangle^{2}+\langle\hat{a}^{\dagger}\rangle^{2}+\langle\hat{b}^{\dagger}\rangle^{2}
\big]\pm(\langle\hat{a}\rangle\langle\hat{b}\rangle+
\langle\hat{a}^{\dagger}\rangle\langle\hat{b}^{\dagger}\rangle)
\notag\\&+\langle\hat{a}\rangle\langle\hat{a}^{\dagger}\rangle
+\langle\hat{a}\rangle\langle\hat{b}^{\dagger}\rangle+
\langle\hat{b}\rangle\langle\hat{a}^{\dagger}\rangle+
\langle\hat{b}\rangle\langle\hat{b}^{\dagger}\rangle\end{align} is
less than one. Following a similar approach as in Section III, it
is possible to see that
\begin{align}\label{dp49}\Delta
c_{\pm}^{2}={\kappa^{2}(\kappa\mp2\varepsilon_{2})+\kappa\varepsilon_{2}^{2}
+4\kappa\varepsilon_{1}^{2}\pm2\kappa\varepsilon_{2}\varepsilon_{1}\over(\kappa\mp\varepsilon_{2})
\big[\kappa(\kappa\mp2\varepsilon_{2})+4\varepsilon_{1}^{2}\big]},\end{align}
which can also be put on the basis of the fact that
$\varepsilon_{2}=\kappa/2$ for $\varepsilon_{1}\ne0$ in the form
\begin{align}\label{dp50}\Delta
c_{\pm}^{2}={\kappa^{2}(5\mp4)+4\varepsilon_{1}(4\varepsilon_{1}\pm\kappa)
\over2(2\mp1)(\kappa^{2}(1\mp1)+4\varepsilon_{1}^{2})}.\end{align}
\begin{figure}[hbt]
\centerline{\includegraphics [height=6cm,angle=0]{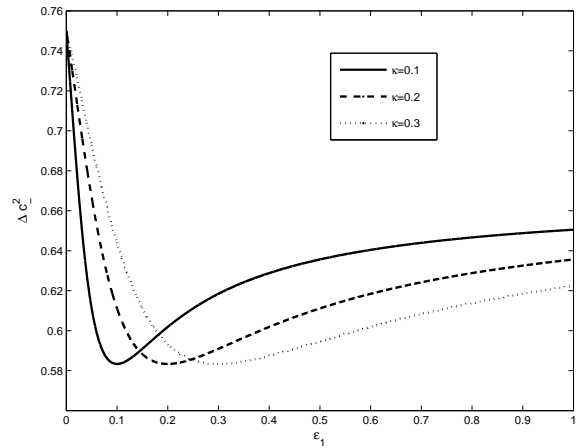} }
\caption { Plots of the minus quadrature variance for the two-mode
cavity radiation ($\Delta c_{-}^{2}$) at steady state
 for different values of $\kappa$.}
\end{figure}

It turns out that the superimposed cavity radiation exhibits a
significant two-mode squeezing for certain values of
$\varepsilon_{1}$ corresponding to each damping constant. As
clearly presented in Fig. 2 the degree of squeezing decreases with
the cavity damping constant for smaller values of
$\varepsilon_{1}$, but it increases for larger values. It is not
difficult to see that a maximum obtainable squeezing occurs
slightly above the threshold value and it is independent of the
cavity damping constant. It is also found that a two-mode maximum
squeezing of about 42\% occurs at different values of
$\varepsilon_{1}$ for different damping constants. The value of
$\varepsilon_{1}$ for which a maximum squeezing occurs increases
with the cavity damping constant. Comparing the results given in
Figs. 1 and 2 shows that two-mode squeezing exits even for values
of $\varepsilon_{1}$ for which there is no entanglement. It goes
without saying that the differences in the entanglement and
squeezing properties are essentially related to the correlations
that lead to these phenomena. That is, the correlations between
similar states of a radiation like $\langle\hat{a}^{2}\rangle$ and
$\langle\hat{b}^{2}\rangle$ contribute to the squeezing, but not
to the entanglement, since entanglement requires two different
states of radiation to be correlated.

\section{Mean of the photon number and intensity difference}

The mean photon number of the cavity radiation corresponding to
the superposition of the two modes can be expressed as
\begin{align}\label{dp51}\bar{n}=\langle\hat{c}^{\dagger}(t)\hat{c}(t)\rangle,\end{align}
where $\hat{c}(t)$ is the annihilation operator defined by Eq.
\eqref{dp43}. Hence it is not difficult to see that
\begin{align}\label{dp52}\bar{n}&={1\over2}\big[\langle\hat{a}^{\dagger}(t)\hat{a}(t)\rangle
+\langle\hat{b}^{\dagger}(t)\hat{b}(t)\rangle\notag\\&+\langle\hat{a}^{\dagger}(t)\hat{b}(t)\rangle
+\langle\hat{b}^{\dagger}(t)\hat{a}(t)\rangle\big].\end{align} In
order to obtain the mean photon number, we need to determine the
involved cross correlations. To this effect, making use of Eqs.
\eqref{dp09}, \eqref{dp10}, \eqref{dp16}, \eqref{dp17},
\eqref{dp18}, \eqref{dp19}, \eqref{dp20}, \eqref{dp21}, and the
correlations of the Langevin noise forces, we find at steady state
\begin{align}\label{dp53}&\langle\hat{a}^{\dagger}(t)\hat{b}(t)\rangle_{ss}
=\langle\hat{b}^{\dagger}(t)\hat{a}(t)\rangle_{ss}=-{\kappa\varepsilon_{1}\varepsilon_{2}
\over4(\varepsilon^{2}_{2}-4\varepsilon_{1}^{2})}\notag\\&\times\left[{\kappa-\varepsilon_{2}
\over\kappa(\kappa-2\varepsilon_{2})+4\varepsilon_{1}^{2}}
-{\kappa+\varepsilon_{2}
\over\kappa(\kappa+2\varepsilon_{2})+4\varepsilon_{1}^{2}}
+{2\varepsilon_{2}\over\kappa^{2}-\varepsilon_{2}^{2}}\right].\end{align}
Thus, applying Eqs. \eqref{dp35}, \eqref{dp36}, \eqref{dp38},
\eqref{dp39}, \eqref{dp52}, and \eqref{dp53}, we arrive at
\begin{align}\label{dp54}\bar{n}&={1\over2}\big(\alpha+\beta\big)^{2}+
{\kappa(\kappa-\varepsilon_{2})(1+p^{2}+q^{2})
\over8[\kappa(\kappa-2\varepsilon_{2})+4\varepsilon_{1}^{2}]}
\notag\\&+{\kappa(\kappa+\varepsilon_{2})(1+p^{2}+q^{2})
\over8[\kappa(\kappa+2\varepsilon_{2})+4\varepsilon_{1}^{2}]}
+{\kappa^{2}(1-p^{2}-q^{2})\over4(\kappa^{2}-\varepsilon_{2}^{2})}
\notag\\&-{\kappa^{2}(1-p^{2}+q^{2})\over4[\kappa^{2}-(\varepsilon_{2}^{2}
-4\varepsilon_{1}^{2})]}
-{(1+p^{2}-q^{2})\over4}-{\kappa\varepsilon_{1}\varepsilon_{2}
\over4(\varepsilon^{2}_{2}-4\varepsilon_{1}^{2})}\notag\\&\times\left[{\kappa-\varepsilon_{2}
\over\kappa(\kappa-2\varepsilon_{2})+4\varepsilon_{1}^{2}}
-{\kappa+\varepsilon_{2}
\over\kappa(\kappa+2\varepsilon_{2})+4\varepsilon_{1}^{2}}
+{2\varepsilon_{2}\over\kappa^{2}-\varepsilon_{2}^{2}}\right],\end{align}
which can also be put for $\varepsilon_{1}\ne0$ in the form
\begin{align}\label{dp55}\bar{n}&={(2\varepsilon_{1}-\kappa)^{2}-4\lambda^{2}\over8\lambda^{2}}
+{\kappa^{3}(\kappa-2\varepsilon_{1})\over32\varepsilon_{1}^{2}(\kappa^{2}
-16\varepsilon_{1}^{2})}
\notag\\&+{3\kappa^{3}(\kappa+2\varepsilon_{1})\over16(\kappa^{2}-16\varepsilon_{1}^{2})
(\kappa^{2}+2\varepsilon_{1}^{2})}-{2\varepsilon_{1}(\kappa+16\varepsilon_{1})\over3(\kappa^{2}-
16\varepsilon^{2}_{1})}.
\end{align}
\begin{figure}[hbt]
\centerline{\includegraphics [height=6cm,angle=0]{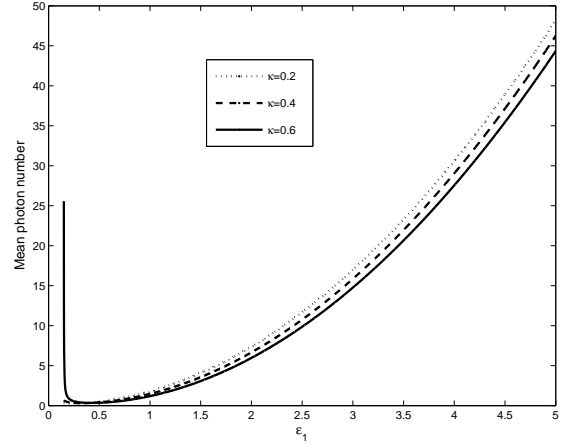} }
\caption {Plots of the mean of the photon number ($\bar{n}$) at
steady state
 for $\lambda=0.5$ and different values of $\kappa$.}
\end{figure}

We see from Eq. \eqref{dp55} that the mean photon number takes
large values where there is no entanglement. Fortunately, the mean
photon number increases with $\varepsilon_{1}$ for larger values
of $\varepsilon_{1}$. Therefore, as we can see from Fig. 3 a
considerably intense entangled two-mode light can be generated
from the driven degenerate parametric oscillator, which we
believed is an encouraging result.

 On the other hand, the
intensity difference  can be defined as
\begin{align}\label{dp56}\Delta \hat{I} =\hat{a}^{\dagger}\hat{a}-\hat{b}^{\dagger}\hat{b}.\end{align}
Upon employing Eqs. \eqref{dp35}, \eqref{dp36}, \eqref{dp38}, and
\eqref{dp39}, the mean of the intensity difference at steady state
turns out to be
\begin{align}\label{dp57}\Delta
I&=\alpha^{2}-\beta^{2}-{\kappa
p\sqrt{\varepsilon_{2}^{2}-4\varepsilon_{1}^{2}}
\over2[\kappa(\kappa-2\varepsilon_{2})+4\varepsilon_{1}^{2}]}
\notag\\&+{\kappa p\sqrt{\varepsilon_{2}^{2}-4\varepsilon_{1}^{2}}
\over2[\kappa(\kappa+2\varepsilon_{2})+4\varepsilon_{1}^{2}]},\end{align}
in which using Eq. \eqref{dp20} along with the fact that
$\varepsilon_{2}=\kappa/2$ for $\varepsilon\ne0$ leads to
\begin{align}\label{dp58}\Delta
I&={4\varepsilon_{1}^{2}-\kappa^{2}\over4\lambda^{2}}
-{\kappa^{2}\over16\varepsilon_{1}^{2}}+{\kappa^{2}\over8(\kappa^{2}+2\varepsilon_{1}^{2})}.\end{align}
\begin{figure}[hbt]
\centerline{\includegraphics [height=6cm,angle=0]{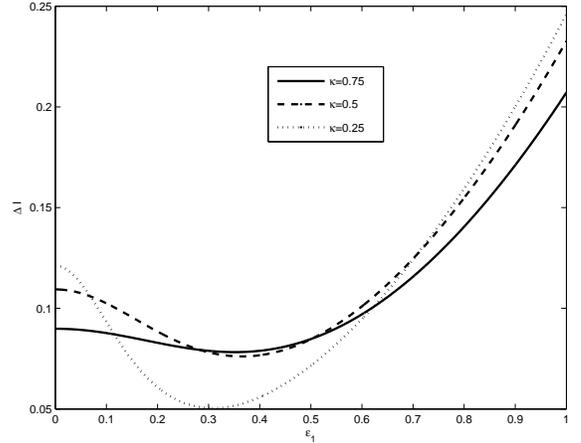} }
\caption {Plots of the mean of the intensity difference ($\Delta
I$) at steady state
 for $\lambda=0.5$ and different values of $\kappa$.}
\end{figure}

According to the results shown in Fig. 4, the mean of the
intensity difference is nonnegative for all values of
$\varepsilon_{1}$. This indicates that, in principle, a degenerate
parametric oscillator can down convert a significant amount of
light falling on it. We notice that the number of down converted
photons available in the cavity as compared to the number of
photons of the driving external coherent radiation depends on the
damping constant, since the fundamental and second-harmonic modes
are assumed to leak through the mirrors in the same way. The mean
intensity difference deceases for smaller values of
$\varepsilon_{1}$, but it increases for larger values. On the
basis of Eq. \eqref{dp15} we note that $\varepsilon_{1}$ increases
with the amplitude of the external coherent radiation provided
that the cavity damping and coupling constants are taken to be
independent of the amplitude of the driving radiation. In view of
this fact we observe that the process of parametric down
conversion decreases with the amplitude of the external coherent
radiation close to the critical point, whereas it will be enhanced
with the external radiation after a particular value of
$\varepsilon_{1}$ which increases with the cavity damping
constant.


\section{Conclusion}

We study the squeezing properties of the two-mode superimposed
radiation and entanglement in the fundamental and second-harmonic
modes of a driven degenerate parametric oscillator coupled to a
two-mode vacuum reservoir. It turns out that the cavity radiation
exhibits a significantly intense two-mode squeezing under certain
conditions pertaining to the rate at which the external coherent
radiation is down converted. A maximum squeezing of about 42\%
occurs slightly above a threshold value for different amplitudes
of the external driving radiation for different damping constants.
On account of Eq. \eqref{dp56} we notice that the mean of the
intensity difference would be positive, provided that more than
33.3\% of the external coherent radiation is down converted.
Therefore, upon comparing Figs. 2 and 4 we see that a two-mode
squeezing exists when greater than 33.3\% of the external coherent
radiation is down converted and maximum squeezing occurs when
$\varepsilon_{1}=\kappa$. We come to understand that although the
squeezing of the subharmonic mode has attracted a great deal of
attention in previous studies, the squeezing of the two-mode
radiation is also equally significant with additional possible
applications.

The strong correlation between the fundamental and second-harmonic
modes due to their harmonic relation leads not only to quadrature
squeezing, but also to entanglement. Unlike the squeezing, the
entanglement increases with the rate at which the external
coherent radiation is down converted. Since the radiation has more
chance to oscillate back and forth in the cavity for smaller
damping constant, the probability that it can be entangled is
better. This must be the reason for the decrement of the
entanglement with the cavity damping constant, which is also true
for the squeezing near threshold. Comparison of Figs. 1 and 4
shows that contrary to the squeezing the entanglement is found to
be better in a region where the down conversion of the external
coherent radiation is enhanced. These differences in squeezing and
entanglement are attributed to the differences in the correlation
leading to these phenomena. It is a well established fact that
entanglement requires a correlation between two different states
of the radiation, a restriction that does not necessarily apply to
harmonically related radiations.


\begin{thebibliography}{1}
\bibitem{1} G. J. Milburn and D. F. Walls, Opt. Commun. {\bf 39},  401 (1981).
\bibitem{2}M. J. Collett and C. W. Gardiner, Phys. Rev.  A {\bf{30}}, 1386  (1984).
\bibitem{3} B. Daniel and K. Fesseha, Opt. Commun. {\bf 151},
384 (1998).
\bibitem{4} L. A. Wu, M. Xiao, and H. J. Kimble, J. Opt. Soc. Am.
B {\bf 4},  1465 (1987).
\bibitem{5} P. Gardiner, R. E. Slusher, B. Yurke, and A. Laporta, Phys. Rev. Lett. {\bf 59},
2153 (1987).
\bibitem{6} K. Fesseha, Opt. Commun. {\bf
156},  145 (1998).
\bibitem{7} J. Anwar and M. S. Zubairy, Phys. Rev.  A {\bf  45},
  1804 (1992).
\bibitem{sint071} S. Tesfa, Nonlinear optics, Quantum optics: Concepts in modern optics (in press).
\bibitem{sint072} S. Tesfa (unpublished).
\bibitem{sintq} S. Tesfa, e-print: quant-ph/0702108.
\bibitem{8} J. Gea-Banacloche,
Phys. Rev. Lett. {\bf 59}, 543 (1987).
\bibitem{9} Y. Zhang, H. Wang, X. Y. Li, J. T. Jing, C. D. Xie, and K. C. Peng,
Phys. Rev. A {\bf{62}},  023813 (2000).
\bibitem{10} C. Simon and D. Bouwmeester, Phys. Rev. Lett. {\bf{91}},  05360 (2003).
\bibitem{11} M. D. Reid and P. Drummond, Phys. Rev. Lett. {\bf{60}},  2731 (1988).
\bibitem{12} M. D. Reid, Phys. Rev. A {\bf{40}},  913 (1989).
\bibitem{13} P. D. Drummond and M. D. Reid, Phys. Rev. A {\bf{41}},  3930 (1990).
\bibitem{14} M. Dance, and M. J. Collett, and D. F. Walls, Phys. Rev. A {\bf{48}},  1532 (1993);
M. K. Olsen and R. J. Horowicz, Opt. Commun. {\bf{168}}, 135
(1999).
\bibitem{15} M. K. Olsen, Phys. Rev. A {\bf{70}},  035801  (2004).
\bibitem{16} O. K. Lim and M. Saffman, Phys. Rev. A {\bf{74}}, 023816   (2006).
\bibitem{17} N. B. Grosse, W.P. Bowen,  K. McKenzie, and P. K. Lam, Phys. Rev. Lett
{\bf{96}},  063601 (2006).
\bibitem{18} A. Einstein, B. Podolsky, and R. Rosen, Phys. Rev. {\bf{47}},  777 (1935).
\bibitem{19} K. Dechoum, P. D. Drummond, and S. Chaturvedi,  Phys. Rev. A {\bf{70}}, 053807 (2004).
\bibitem{lod} P. Lodahl, Phys. Rev. A {\bf{68}},   023806 (2003).
\bibitem{21} L. Gilles,  P. Tombesi, M. San Mingueland, and P.
Garcia-Fernandez, Phys. Rev.   A {\bf 55},  2245 (1997).
\bibitem{22} S. Chaturvedi, K. Dechoum, and P. D. Drummond, Phys. Rev. A {\bf{65}}, 033805 (2002).
\bibitem{prl93063601} J. Fiurasek and N. J. Cerf, Phys. Rev. Lett. {\bf{93}},  063601 (2004).
\bibitem{prl95120502} J. Korbicz, J. I. Cirac, and M. Lewenstein, Phys. Rev. Lett.
{\bf{95}}, 120502 (2005).
\bibitem{pra74} S. Tesfa, Phys. Rev. A  {\bf{74}}, 043816 (2006).
\bibitem{duan} L. M. Duan, G. Giedke, J, I, Cirac, and P. Zoller, Phys. Rev. Lett
{\bf{84}},  2722
 (2000).

\end{thebibliography}
\end{document}